\documentclass{article}
\usepackage[utf8]{inputenc}
\usepackage{amsmath}
\usepackage{cite}
\usepackage{graphicx}

\title{Current Mapping from the Wave Spectrum}
\author{Benjamin K. Smeltzer and Simen \AA. Ellingsen}

\begin{document}

\maketitle


In this chapter we review methods by which near--surface ocean currents can be measured remotely using images of the water surface, as obtained by X-band radar in particular. The presence of a current changes the dispersive behavior of surface waves, so our challenge is to solve the inverse problem: to infer the spatially-varying current from measurements of the wavy surface. Measuring near-surface currents in the ocean is important for a large variety of applications. Examples include oil spill tracking, understanding the transport of micro-plastics, predicting loads on marine structures, fuel optimization for ships, and providing data to inform oceanographic and climate models, among many others\cite{kudryavtsev08,belcher12,laxague17,laxague18,lund18,gangeskar18,zippel17,dalyrmple73}. Many in situ methods for measuring currents experience a broad range of challenges such as time-consuming and expensive deployment and maintenance, and noise in the measurements from wave or platform motions, fouling, or other factors\cite{gangeskar18}. Remote sensing of currents is an attractive alternative to in situ measurements, as currents can be mapped over a finite areal extent simultaneously, and the deployment and retrieval of sensors is not needed.

\begin{figure}[h]
\centering \includegraphics[scale=0.7]{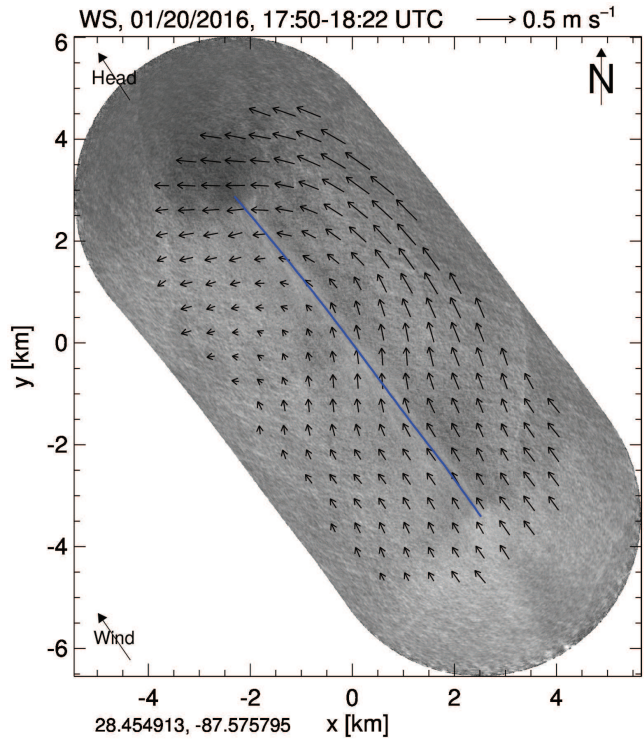}
\caption{An example of a current vector field obtained from X-band radar. Figure taken from Reference \cite{lundRep}.}
\label{fig:map}
\end{figure}

We here examine how remote sensing of currents is achieved in practice by analyzing the wave spectrum, as may be measured by X-band radar, cf. e.g. \cite{young85,lund15,campana17}. A set of consecutive backscatter images recorded as a function of time is Fourier-transformed to produce the spectrum, which gives information concerning the propagation of waves whose dispersion is altered by currents. X-band radar images measure the wave field over multiple square kilometers, and analyzing various spatial subsets of the images allows a map of the spatial variation of the currents to be reconstructed, cf. e.g. \cite{hessner14,gangeskar18,lund18}. An example map where arrows show surface currents calculated from wave spectra is shown in Figure \ref{fig:map}, demonstrating the areal coverage and spatial resolution that may be achieved. Thus, with a single sequence of radar images, the currents can be mapped over a large area simultaneously, where achieving the same degree of spatial coverage using in situ point sensors would require a large-scale deployment effort. 

We describe herein methods for reconstructing currents from a measured wave spectrum. It is worth mentioning that although the methods presented in this chapter are described in the context of marine radar, they may be readily applied to wave spectra measured by other means as well, e.g.\cite{horstmann17,micha17,laxague17}. The starting point for this chapter is thus the directional frequency-wavenumber spectrum, which may in theory be obtained by a variety of means.


\section{Wave propagation atop background currents}
We briefly introduce the dispersion relation for waves traveling atop background currents, which governs wave propagation and is the basis for determining currents from measurements of the wave spectrum. The dispersion relation describes the relationship between the wave frequency and wavenumber, which for quiescent waters reads:

\begin{equation}
\omega_0(\mathbf{k}) = \sqrt{gk\tanh kh},
\label{eq:dr0}
\end{equation}
where $\omega_0$ is the wave angular frequency, $k = |\mathbf{k}|$, $g$ is the acceleration due to gravity, and $h$ the water depth. Surface tension has been neglected, which is a valid assumption for waves measured by marine radar. In cases where the water depth is greater than roughly half the relevant wavelength, $\tanh kh \approx 1$ and the deep water limit may be used ($\omega_0 = \sqrt{gk}$). Throughout the chapter we use $\omega_0$ to denote the wave frequency in quiescent waters, with the implicit understanding that finite depth must be taken into account when relevant.
\begin{figure}[h]
\centering \includegraphics[scale=0.6]{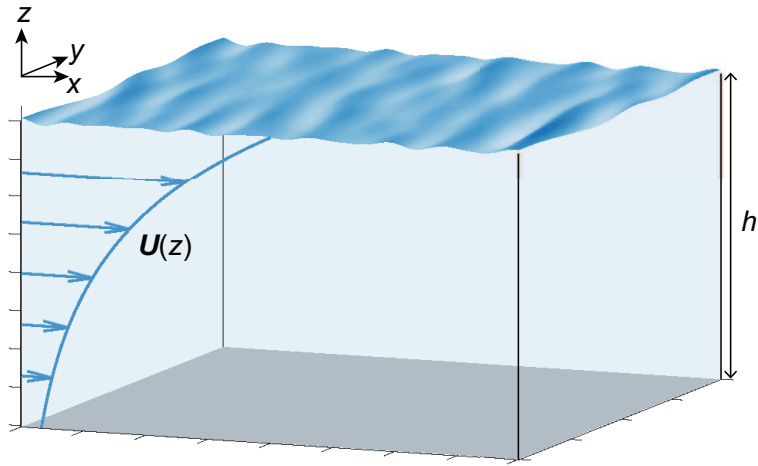}
\caption{The basic geometry and coordinate system used in this chapter. A general form of the current profile $\mathbf{U}(z)$ is shown; in part of the chapter a depth-uniform profile is assumed.}
\label{fig:geom}
\end{figure}
We consider a background current moving in the horizontal plane: $\mathbf{U}(\mathbf{r}) = [U_x(\mathbf{r}), U_y(\mathbf{r})]$, where $\mathbf{r} = [x,y,z]$. The geometry and coordinate system is shown in Figure \ref{fig:geom}. It is noted that some qualitative information concerning currents moving the vertical direction may be obtained by consider the spatial variation of the horizontal current components together with the continuity equation (e.g. to identify regions of up-welling or down-welling), but herein we neglect any discussion of vertical currents. The entire spatial domain of a marine radar image spans multiple km, within which the strength and direction of currents may vary significantly (see Figure \ref{fig:map}). To proceed, we consider a small subset spatial window of the domain, in which it will be assumed that the variation of $\mathbf{U}$ in the $x$ and $y$ directions is negligible, such that $\mathbf{U} = \mathbf{U}(z)$ as sketched in Figure \ref{fig:geom}. This assumption drastically simplifies the analysis of the wave spectrum as refraction effects may be neglected. The subset spatial window we consider corresponds in Figure \ref{fig:map} to a local spatial extent centered on the location of one of current vectors. By reconstructing the currents within each individual subset spatial window, a full map of the horizontal variation of the currents can be achieved at the resolution of the window size. Smaller window sizes thus give higher resolution in the reconstructed current field, but at the cost of lower spectral resolution which may decrease the accuracy of the sensed currents (discussed below in the next section).

In addition to varying in the horizontal plane, $\mathbf{U}$ may also vary with depth, particularly in the vicinity of the surface. This variation is often neglected as it complicates the analysis, and the current is therefore often assumed to be depth-uniform, equal to its surface value at all $z$. We use this assumption for the first part of the chapter in section \ref{sec:alg}. However, the depth dependence may also be extracted from the wave spectrum, and we describe methods to achieve this later in the chapter in section \ref{sec:depth}. 

\begin{figure}[h]
\centering \includegraphics[scale=0.9]{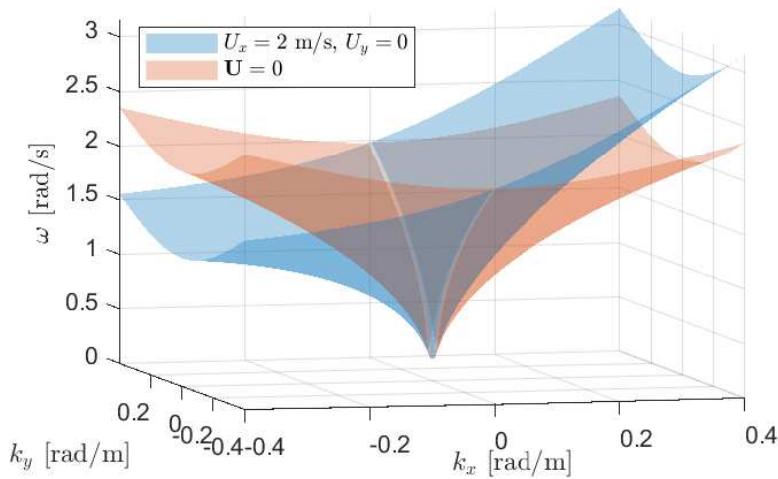}
\caption{The dispersion surface $\omega_{\mathrm{DR}}$ for the case of a current of 2 m/s in the $x$-direction, and in quiescent waters.}
\label{fig:dr}
\end{figure}

For now, we assume the currents are constant in all spatial dimensions, in which case the dispersion relation may be written as:

\begin{equation}
\omega_{\mathrm{DR}}(\mathbf{k}) = \omega_0 + \mathbf{k}\cdot\mathbf{U}.
\label{eq:dr}
\end{equation} 
Examples of the dispersion surface $\omega_{\mathrm{DR}}(\mathbf{k})$ for the case of $\mathbf{U} = (2 \mathrm{\ m/s}, 0)$ and $\mathbf{U} = 0$ are shown in Figure \ref{fig:dr}.

It is noted that $\omega_{\mathrm{DR}}$ and the velocity $\mathbf{U}$ are defined in the reference frame of the radar system, a distinction relevant when mounted on a moving platform such as a ship, and $\mathbf{U}$ in Eq. \ref{eq:dr} is often termed the ``velocity of encounter.'' In such cases, accurate geographic registration of the radar images is important \cite{lund15a,mccann18}. ``Calibration'' methods are presented in References \cite{lund15a} and \cite{mccann18} for handling errors in ship heading measurements as well as other error sources. 


\section{Appearance of the linear dispersion relation in the spectrum}
When taking a fast Fourier transform (FFT) of a set of radar backscatter images of the sea surface, the resulting spectrum includes components spanning frequencies and wavenumbers up to the Nyquist frequency in both time and space. The spectrum is thus a function of three variables: wavenumber components $k_x$ and $k_y$, and frequency $\omega$. From the wave dispersion relation (\ref{eq:dr}), we see that only certain combinations of ($k_x$, $k_y$, $\omega$) or ``triplets'' are allowed by the physics: for a particular wavevector, the linear dispersion relation gives a unique frequency component. Thus, in examining the spectrum obtained by FFT, the greatest signal is expected to lie in frequency-wavenumber triplets that satisfy the dispersion relation. 

The spectrum is typically defined as:

\begin{equation}
P(k_x,k_y,\omega) = |\mathrm{FFT}\{I(x,y,t)\}|^2,
\label{eq:spec}
\end{equation}
where $I$ is the radar backscatter image intensity. Given that $I$ is a real quantity and $P$ the square magnitude of the Fourier transform, a point symmetry about the origin $P(0,0,0)$ applies:

\begin{equation}
P(k_x,k_y,\omega) = P(-k_x,-k_y,-\omega).
\label{eq:ps}
\end{equation}
Due to the symmetry expressed in Eq. \ref{eq:ps}, there are two frequencies $\omega_\pm$ in practice associated with the dispersion relation for a particular wavevector: $\omega_+(\mathbf{k}) = \omega_{\mathrm{DR}}(\mathbf{k})$, and $\omega_-(\mathbf{k}) = -\omega_{\mathrm{DR}}(-\mathbf{k})$. Given appropriate handling of this symmetry in algorithms, in practice only half the spectrum is necessary for analysis which reduces computational demands.

It is noted that the spectrum $P$ may be normalized by a function $N(\omega,k)$ defining the background noise, to produce a signal-to-noise (SNR) frequency-wavenumber spectrum, e.g. \cite{lund15}. The noise spectrum is typically greater at lower frequencies and wavenumbers, and the normalization results in a SNR spectrum where the values of the peaks are more uniform over frequencies and directions. We use the spectrum $P$ in this chapter with the understanding that it may correspond to that precisely defined by (\ref{eq:spec}) or the SNR spectrum.


\subsection{Practical considerations}
\label{sec:prac}
Though the maximum signals in the measured wave spectrum are expected to lie on the linear dispersion relation surface, practical analysis of the spectrum is complicated by several factors. At root, some of these factors arise from the finite sampling in space and time of the radar images, as well as their finite temporal and spatial extent. Other factors are due to imperfections of the radar imaging of waves as well as the underlying physics of the waves themselves, which result in  additional signatures in the spectrum than simply the linear dispersion relation. We outline several of these practical issues in this section.
\subsubsection{Spectral resolution}
\label{sec:specres}
The finite extent of the recorded radar images of the waves in space and time determines the resolution of the wave spectrum in wavenumber and frequency. For spatial extent $L_x$, $L_y$, and $T$ in the spatial and temporal dimensions respectively, the extent of one pixel in the spectral domain is $\frac{2\pi}{L_x} \times \frac{2\pi}{L_y} \times \frac{2\pi}{T}$. The finite spectral resolution directly affects the precision with which the location of the energy peaks corresponding to the dispersion relation may be determined. It is thus desirable to maximize the spectral resolution especially when considering lower wavenumbers whose frequencies are less sensitive to currents due to the $\mathbf{k}$-proportionality in (\ref{eq:dr}). In the spatial domain, this entails larger spatial windows which then decreases the spatial resolution of the reconstructed current field map. In the temporal domain, $T$ may be increased by recording more images (provided the currents don't vary appreciably during this duration), which increases the size of the dataset and slows processing. In practice $T$ may be limited in cases where the radar system is mounted on a moving platform.

\subsubsection{Harmonics}
Harmonics of the linear dispersion relation arise in the spectrum from multiple sources, cf. e.g. \cite{senet01}. First, there is the nonlinearity of the radar imaging system: the mapping between wave height and radar signal intensity is not entirely linear. This results in components in the spectrum at integer multiples of the frequency-wavenumber combinations lying along the linear dispersion relation

\begin{equation}
S_p(\mathbf{k}) = \pm(\sqrt{p+1}) \omega_0(\mathbf{k}) + \mathbf{k}\cdot\mathbf{U},
\label{eq:harm}
\end{equation}
with $p$ a positive integer $\geq 0$. Examples of $S_p$ are shown in Figure \ref{fig:aliasing}a) for $p = 0,1,2$.

There is in addition non-linearities from the surface waves themselves, which do not form perfect sinusoidal surface profiles when finite in amplitude. Their harmonic signatures in the spectrum may also be described by Eq. \ref{eq:harm}, making them in practice difficult to distinguish from imaging nonlinearities. For most realistic cases, the wave nonlinearity spectral structure is likely smaller than that due to imaging non-linearities, except perhaps under extreme sea states.

\subsubsection{Aliasing}
\label{sec:alias}
Aliasing is an artifact of under-sampling either in time or space. For spectra obtained by marine radar, temporal under-sampling is often more relevant given the relatively slow rotation rate of a radar antenna, as well as the possibility for large velocities of encounter when mounted on a moving platform. Given a sampling frequency $f_S$, all frequencies $|\omega| > \pi f_S$ will be under-sampled. The frequency $\pi f_S\equiv\omega_N$ is known as the Nyquist frequency, representing the largest frequency (in magnitude) that will be adequately sampled.

To understand the signatures in the wave spectrum from aliasing, we consider two spectra $P_1$ and $P_2$, consisting only of monochromatic waves at a frequency $\omega_1$ in the case of $P_1 $ and $\omega_2 = \omega_1 +2n\omega_N$ for $P_2$, where $n$ is an integer. Due to the finite sampling frequency, it can be shown that $P_1 = P_2$, thus wave components with frequencies differing by $2n\omega_N$ are indistinguishable in the spectrum. An under-sampled wave frequency $\omega_{\mathrm{aliased}}$ will appear in the spectrum at a frequency 
\begin{equation}
\omega = \omega_{\mathrm{aliased}} + 2 n \omega_N,
\label{eq:alias}
\end{equation}
with $n$ such that $\omega$ lies within the interval $[-\omega_N, \omega_N]$. 

\begin{figure}[h]
\centering \includegraphics[scale=0.85]{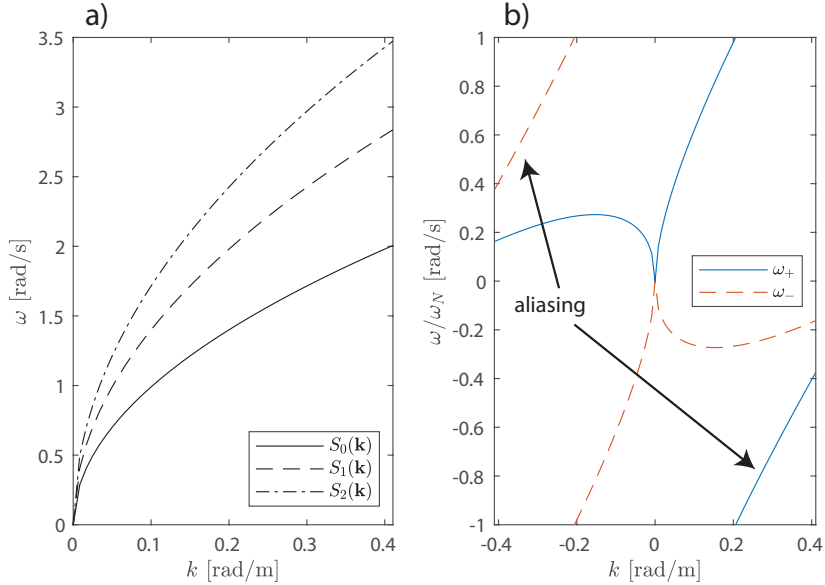}
\caption{a) Equation \ref{eq:harm} for the fundamental harmonic ($p=0$) and first two harmonics. b) Frequencies $\omega_\pm + 2n\omega_N$ are plotted for the case of a strong current as a function wavenumber component $k$ aligned with the current. The $2\omega_N$-periodicity has been included to illustrate the effects of aliasing in the spectrum.}
\label{fig:aliasing}
\end{figure}

When examining the wave spectrum, aliasing manifests itself as extra artifacts of energy located away from the linear dispersion relation. An illustrative example is shown in Figure \ref{fig:aliasing}b). Frequencies $\omega_pm + 2n\omega_N$ are plotted for the case of a strong current aligned with the waves ($k$ in this case is the component of $\mathbf{k}$ along the direction of the current). Considering $\omega_+$, the wave frequency is greater than the Nyquist frequency for wavenumbers above $\sim 0.2$ rad$/$m resulting in aliased components appearing at negative frequencies corresponding to the $n = -1$ branch in (\ref{eq:alias}). The $\omega_-$ curve displays an analogous characteristic. Equation \ref{eq:alias} and the symmetry relation (\ref{eq:ps}) may be used to perform de-aliasing to determine the true under-sampled wave frequency $\omega_{\mathrm{aliased}}$ to which the spectral signal corresponds \cite{senet01}. It is noted that the harmonics in (\ref{eq:harm}) also will be aliased resulting in a more complicated spectrum to interpret, especially in the presence of an unknown current velocity.

\section{Extracting currents from the spectrum}
We have seen thus far how the wave dispersion relation, indicative of wave propagation, manifests itself as peaks in the measured wave spectrum. Location of the peaks thus allows a measurement of the dispersion relation. Assuming a depth uniform flow, the wave dispersion relation is described by (\ref{eq:dr}), and the goal is to determine the unknown current velocity $\mathbf{U}$. We now examine algorithms to extract the current by analyzing the peaks in the spectrum.
\label{sec:alg}
\subsection{Least squares method}
\begin{figure}[h]
\centering \includegraphics[scale=0.85]{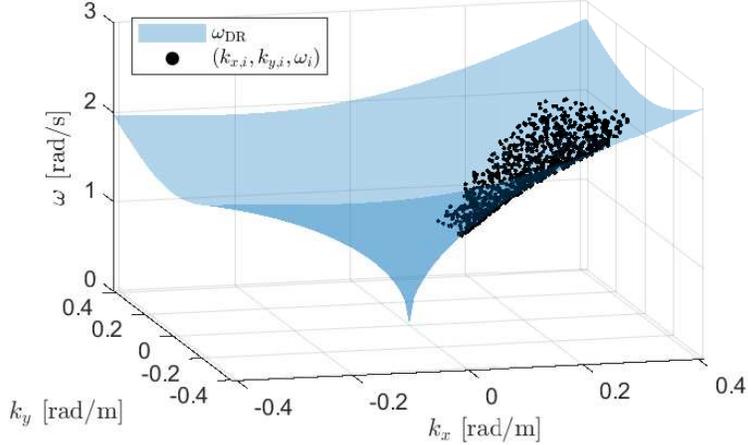}
\caption{Illustration of wavenumber-frequency triplets (black circles) extracted from peaks in the spectrum corresponding to the linear dispersion relation (blue surface), using mock data.}
\label{fig:triplets}
\end{figure}

A least squares (LS) method was proposed for extracting currents from the wave spectrum by Young \& Rosenthal\cite{young85}. Wavenumber-frequency triplets $(k_{x,i},k_{y,i},\omega_i)$ corresponding to the linear dispersion relation are identified as values of the spectrum above a certain threshold $C_1$ of the maximum value satisfying:

\begin{equation}
P(k_{x,i},k_{y,i},\omega_i)\geq C_1\mathrm{max}\{P(k_x,k_y,\omega)\}.
\label{eq:thresh}
\end{equation}
An example of a set of triplets is shown in Figure \ref{fig:triplets} as the black circles. Then, an error parameter is defined as: 
\begin{equation}
Q(\mathbf{U}) = \sum_{i=1}^{N_1}(\omega_i - \omega_0(\mathbf{k}_i) - \mathbf{k}_i\cdot\mathbf{U})^2,
\label{eq:ls}
\end{equation} 
where the summation is over all $N_1$ triplets selected by the threshold criteria (\ref{eq:thresh}). $Q$ is then minimized with respect to $\mathbf{U}$ to obtain the unknown current components.

The choice of the threshold value is a tradeoff between two factors. On the one hand, the precision of the least squares fit generally improves with an increasing number of selected triplets $N_1$, achieved with a lower threshold value. On the other hand, as the threshold decreases, other signatures in the spectrum such as harmonics, aliasing, and noise will begin to be selected which corrupts the fit. Ideally, the optimal threshold value would thus be as small as possible while avoiding artifacts and noise. In practice a value of 0.2 is common\cite{huang12}. Several related least squares algorithms have been proposed, such as in Reference \cite{gangeskar02}.

Referring to section \ref{sec:prac}, the signatures in the spectrum from harmonics and aliasing also contain information about the currents which may be used in the fit if we modify the least squares fitting algorithm to include them. The advantage is a greater number of triplets and increased precision. Senet \textit{et al.} \cite{senet01} proposed such an algorithm that is iterative in nature, and termed the iterative least squares (ILS) method. First, the conventional LS method (\ref{eq:ls}) is used to obtain a coarse guess of the current. Second, the value of the current is used to select triplets in the spectrum corresponding to both the fundamental and higher harmonics using a much smaller threshold parameter $C_2$, performing de-aliasing as outlined in section \ref{sec:alias} using (\ref{eq:alias}), (\ref{eq:harm}), and (\ref{eq:ps}). Further practical details concerning an approach to de-aliasing are given in Reference \cite{senet01}. De-aliasing approaches arrive at a function $S_p(\mathbf{k})$ expressing the de-aliased frequencies at harmonic $p$ using the current velocity of iteration $j$. A corresponding error parameter is then defined analogous to (\ref{eq:ls}):

\begin{equation}
Q(\mathbf{U}) = \sum_{i=1}^{N_1}(\omega_i - S_p(\mathbf{k}_i) - \mathbf{k}_i\cdot\mathbf{U})^2,
\label{eq:ils}
\end{equation}
which is then minimized to find an updated current velocity. The process is then repeated, with the de-aliasing being performed with the updated current velocity. The algorithm may be set to run for a fixed number of iterations, or until satisfactory convergence is reached.


\subsection{Normalized scalar product method}
A conceptually related algorithm to the least squares involves the maximization of the normalized scalar product (NSP) between the spectrum and a characteristic function $G$ defining the dispersion relation shell\cite{serafino10,huang12,huang16}

\begin{equation}
G(\mathbf{k},\omega,\mathbf{U}) = 
\begin{cases}
1, & \text{if } |\omega_0(k) + \mathbf{k}\cdot\mathbf{U} - \omega|\leq \Delta\omega/2\\
0,   & \text{otherwise}
\end{cases},
\label{eq:g}
\end{equation}
where $\Delta\omega$ is the frequency resolution of the spectrum (see section \ref{sec:specres}). The characteristic function defines the dispersion shell having a full width of $\Delta\omega$ in frequency for each wavevector. In theory, with the correct current velocity $\mathbf{U}$, the characteristic function should overlap with the peaks in the measured spectrum. The normalized scalar product expresses this overlap and is typically defined as:

\begin{equation}
V(\mathbf{U}) = \frac{\langle|F_I(\mathbf{k},\omega)|,G(\mathbf{k},\omega,\mathbf{U})\rangle}{\sqrt{P_FP_G}},
\label{eq:nsp}
\end{equation}
where $F_I = \sqrt{P}$ and $P_F$ and $P_G$ are the power of $F_I$ and $G$ respectively. The NSP method maximizes $V$ by searching over appropriate ranges of current velocities $\mathbf{U}$. Compared with the LS and ILS methods, the NSP method does not involve the choice of a threshold parameter, reducing the user-input parameters that may affect the results. A potential disadvantage with the NSP method is that the maximized metric $V$ tends to weights by lower wavenumbers which typically have the largest energy in the spectrum and are less sensitive to currents than higher wavenumbers.

Another potential drawback of the NSP method is the computational cost, as (\ref{eq:nsp}) must be evaluated for many values of the current velocity in the search process. To reduce computational demands, a variable search range has been used\cite{huang12}. Typically, an initial wide search range with coarse velocity resolution is used to produce a rough estimate of the current velocity. Then, a narrower range with finer resolution is used centered on the initial estimate to find a higher precision current velocity. The latter step may be repeated to give successively higher precision if desired.

\subsection{Polar current shell method}
An algorithm which transforms the wavenumber plane into polar coordinates to then determine the current is known as the polar current shell (PCS) method\cite{shen15,huang16}. The first step is analogous to the LS algorithm: wavenumber triplets in corresponding to the linear dispersion relation are identified in the spectrum by locating peaks in the frequency for a given wavevector. The wavenumber coordinates of the triplets are then converted into polar coordinates $(k,\theta)$, where $k_x = k\cos\theta$ and $k_y = k\sin\theta$ with $\theta$ being the angle in the $x,y$ plane from the positive $x$ axis. It is noted that (\ref{eq:dr}) expressed in polar coordinates may take the form:

\begin{equation}
\omega_{\mathrm{DR}}(k,\theta) = \omega_0(k) + kU\cos(\theta-\theta_U),
\label{eq:drPolar}
\end{equation} 
where $U = |\mathbf{U}|$ and $\theta_{U}$ is the angle reflecting the direction of the current. We see from inspection of (\ref{eq:drPolar}) that when $\omega_{\mathrm{DR}}$ is evaluated at a particular wavenumber as a function of $\theta$, i.e. along the azimuthal direction, the frequency is a sum of a $\theta$-independent component $\omega_0$ and an oscillating component $kU\cos(\theta-\theta_U)$. Considering the latter, the oscillation amplitude is proportional to the strength of the current, while the phase of the oscillation determines the current direction. 

The PCS algorithm analyzes the $\theta$-dependence of the frequency at a particular wavenumber to find the current from the set of triplets. A least squares fit is performed to extract the current magnitude and direction from a subset of triplets where the wavenumber is constant. The result is a set of current velocities over a range of wavenumbers which are then averaged to give a single vector. Further practical details on the implementation of the PCS are given in References \cite{shen15} and \cite{huang16}.

A related algorithm which is essentially a modified implementation of the PCS was developed by Smeltzer \textit{et al}. \cite{smeltzer19}. Instead of extracting triplets from the spectrum, a NSP was defined using a characteristic function expressing the azimuthal dependence of (\ref{eq:drPolar}) for constant wavenumber. The characteristic function may be expressed (slightly modified from Reference \cite{smeltzer19})

\begin{equation}
G_i(\omega,\theta,U_i,\theta_{U,i}) = \mathrm{exp}\left[\frac{(\omega-\omega_{\mathrm{DR}}(k_i,\theta))^2}{4a}\right],
\label{eq:nspPEDM}
\end{equation}
where $a$ is a frequency width parameter (typically set to a value on the order of the frequency resolution), and the subscript $i$ denotes a discrete wavenumber in the spectrum at which the NSP is evaluated. The dependence of the right hand side on $U_i$ and $\theta_{U,i}$ is implicitly included in $\omega_{\mathrm{DR}}$. Using (\ref{eq:nspPEDM}) the NSP was maximized analogous to (\ref{eq:nsp}) for each separate wavenumber.

\subsection{Algorithm comparison}
References \cite{huang12,huang16} have compared the performance of some of the algorithms described above and largely found similar accuracy. The ILS and NSP methods show comparable accuracy, with ILS being an improvement over the LS. Comparison to the PCS algorithm also showed comparable accuracy in another study \cite{huang16}. The authors however note that the comparison was performed for low velocities of encounter with minimal aliasing. One motivation for developing the NSP \cite{serafino10} was to overcome the problem of increasing errors of the LS and ILS methods for large velocities of encounter, so it is possible the NSP is a better choice is such situations. Referring to the similar performance of the various algorithms the authors in Reference \cite{huang16} conclude that: ``This implies that the technology of current measurement using
X-band marine radar has become sufficiently mature and the emphasis perhaps may be legitimately shifted from research methodology toward applications.''


\section{Reconstructing depth-dependent flows}
\label{sec:depth}
So far in this chapter we have made the assumption that the currents extracted from the wave spectrum are uniform in depth. This assumption drastically simplifies the analysis of the wave spectrum.  However, in some realistic situations currents have significant variation with depth, such as created by wind forcing or a river plume for example. The wave dispersion may be approximated by a different relation:

\begin{equation}
\omega_{\mathrm{DR}}(\mathbf{k}) = \omega_0(\mathbf{k}) + \mathbf{k}\cdot\tilde{\mathbf{c}}(k),
\label{eq:DRshear}
\end{equation}
where $\tilde{\mathbf{c}}$ is a wavenumber-dependent Doppler shift velocity from the background current. The wavenumber-dependence is the key difference to (\ref{eq:dr}). The Doppler shift velocity is a weighted average of the current as a function of depth, approximated as\cite{stewart74}:

\begin{equation}
\tilde{\mathbf{c}}(k) = 2k\int_{-\infty}^0\mathbf{U}(z)\mathrm{e}^{2kz}\mathrm{d}z
\label{eq:SJ}
\end{equation}
in deep water, and in finite water depth as\cite{skop87,kirby89}

\begin{equation}
\tilde{\mathbf{c}}(k) = \frac{2k}{\sinh(2kh)}\int_{-h}^0\mathbf{U}(z)\cosh[2k(h+z)]\mathrm{d}z.
\label{eq:KC}
\end{equation}
As may be noticed from inspecting both expressions, shorter wavelengths are thus influenced by currents in close vicinity to the surface, while longer wavelengths are influenced by currents at greater depths. 

Equations (\ref{eq:SJ}) and (\ref{eq:KC}) reveal the nature of the error one makes by making the common assumption of a depth--uniform current as in section \ref{sec:alg} when in fact $\mathbf{U}(z)$ has a significant variation with depth. What is measured is then not the surface current as commonly reported, but rather a weighted average of the current in the topmost part of the water column. The weighting factor $\exp(2kz)$ means the influence nearest the surface is strongest, and rapidly decreases at a rate which, importantly, depends strongly on the wavelength. The current at depths down to a quarter of the wavelength or so is significant. In effect the current is thus measured at some depth beneath the surface, where $\mathbf{U}(z)$ equals the measured velocity (since $\mathbf{U}(z)$ is assumed to be a smooth function such a depth exists). It is not straightforward to surmise the exact depth, however, since the surface current as found in section \ref{sec:alg} is determined from a spectrum of different wavelengths, and the form of $\mathbf{U}(z)$ is \emph{a priori} unknown. This point becomes particularly important when comparing radar--derived currents to in situ measurements which are typically point measurements at a given depth. When $\mathbf{U}(z)$ is approximately constant the two may be directly compared, but in general the comparison is more complicated. 


Starting again from the measured wave spectrum, reconstructing depth-dependent currents involves two general steps. First, Doppler shift velocities are extracted from the spectrum at a range of wavenumber values. Second, the set of Doppler shifts are used to estimate the unknown profile $\mathbf{U}(z)$. The first step is similar to the methods described in section \ref{sec:alg}. The difference is that while a single current velocity was derived from the spectrum spanning all wavenumbers in section \ref{sec:alg}, multiple velocities representing the second term in (\ref{eq:DRshear}) are found, each corresponding to a unique wavenumber. In practice, this is accomplished by only considering a narrow range of wavenumbers at a time (a bin), and then using one of the methods described in section \ref{sec:alg} to find a velocity that is assigned to the center-wavenumber value of the particular bin. The result is a set of velocities each corresponding to a discrete wavenumber value. Example results are shown in Figure 6 of Reference \cite{lund15} and Figure 9-10 of Reference \cite{lund20}. Small wavenumbers to a larger extent represent the current at greater depths compared to high wavenumbers, which can be seen by inspecting the integrand in (\ref{eq:SJ}).


The second step uses an inversion method to find the depth profile from the Doppler shift velocities.  As described above, the Doppler shift velocities reflect a weighted average of the current profile over different depth ranges depending on the wavenumber. Thus, the best performance is obtained when there are Doppler shifts for a wide range of wavenumbers. In addition, the depth range over which the currents can be reconstructed is also dependent on the wavenumber range: the smallest wavenumbers influence the greatest depth at which the waves ``see'' the flow. In addition, the process of determining the unknown current profile from the Doppler shift velocities is an ill-posed problem mathematically. The resulting current profile is not necessarily mathematically unique, and furthermore, errors in the Doppler shifts tend to be amplified in the inversion process. Because of these challenges, many inversion methods use a priori assumptions and constraints on the functional form of the current profile to produce realistic estimates. In this section we describe several inversion methods that have been used to reconstruct a depth profile estimate from a set of Doppler shift velocities measured at discrete wavenumbers.

%

\subsection{Effective depth method}
Assuming a profile where that current strength varies linearly with depth $\mathbf{U}(z) = \mathbf{U}^\prime z + \mathbf{U_0}$, with $\mathbf{U}^\prime$ the shear-strength (vorticity) and $\mathbf{U_0}$ the surface current. Assuming deep water and using (\ref{eq:SJ}), the Doppler shifts can be expressed as\cite{stewart74}

\begin{equation}
\tilde{\mathbf{c}}(k) = -\frac{\mathbf{U}^\prime}{2k} + \mathbf{U_0} = \mathbf{U}(z = -(2k)^{-1}).
\label{eq:edmlin}
\end{equation}
We see from inspecting (\ref{eq:edmlin}) that the Doppler shifts are equal to the current profile at a depth $Z_{\mathrm{eff}} = -(2k)^{-1}$. This effective depth is roughly 8\% of the wavelength. 

Similarly for a logarithmic profile, $\mathbf{U}(z) = \mathbf{U_0} - \frac{\mathbf{u}^*}{\kappa}\log\frac{z}{z_0}$, where $\mathbf{u}^*$ is the friction velocity, $\kappa$ the von K\'{a}rm\'{a}n constant, and $z_0$ the roughness length. The above parameters characterize a turbulent boundary layer which has been hypothesized to be a reasonable model of a wind-driven shear flow near the water surface\cite{wu75}. Again using (\ref{eq:SJ}), the Doppler shifts are evaluated as\cite{plant80}:

\begin{equation}
\tilde{\mathbf{c}}(k) \approx \mathbf{U_0} - \frac{\mathbf{u}^*}{\kappa}\log\left(\frac{1}{2kr}\frac{1}{z_0}\right) = \mathbf{U}(z = -(3.56k)^{-1}),
\label{eq:edmlog}
\end{equation}
with $r = 1.78$. We see, as with (\ref{eq:edmlin}), that the Doppler shifts are equal to the current profile at a particular depth $Z_{\mathrm{eff}} = -(3.56k)^{-1}$, or 4.5 \% of the wavelength. The only difference from a linear profile is the proportionality of the inverse wavenumber-dependence, where Doppler shifts are mapped to shallower depths relative to a linear current profile.

\begin{figure}[h]
\centering \includegraphics[scale=0.80]{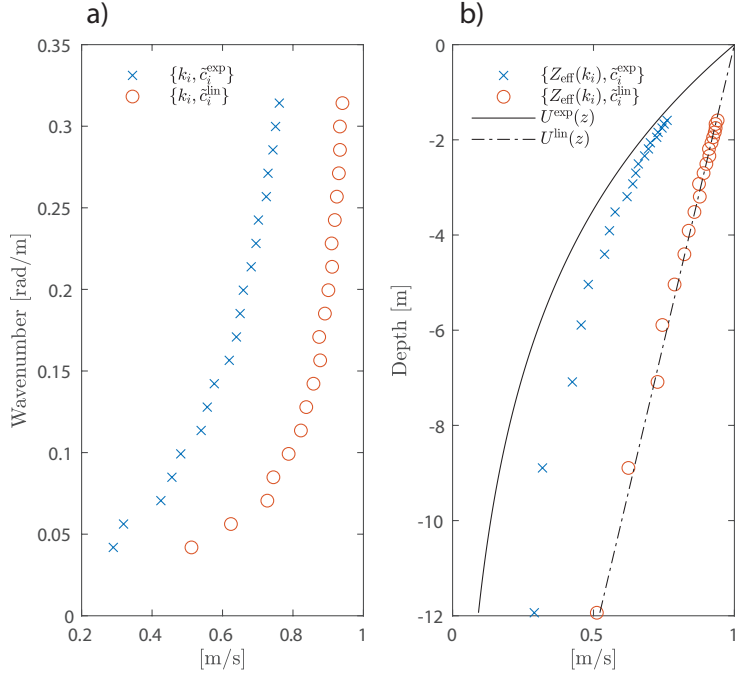}
\caption{a) Doppler shift velocities in the direction of the current as a function of wavenumber for an exponential profile and linear profile (superscripts `exp' and `lin' respectively). b) Mapped velocities assuming a linear current profile, with the true current profiles shown as the solid (exponential) and dashed-dotted (linear) curves for comparison. See text for the parameters defining the current profiles.}
\label{fig:edm}
\end{figure}

This process of mapping Doppler shift velocities to depth we term the effective depth method (EDM). The EDM has been used extensively in the past with both the linear and logarithmic profile assumptions to produce estimates of the depth profile\cite{fernandez96,laxague17,laxague18,lund15,stewart74,teague01}. The main advantage is the simplicity of implementation: once the Doppler shift velocities have been extracted from the spectrum, all that remains is the mapping step using $Z_{\mathrm{eff}}(k)$ to produce current velocities at various depths. If a smooth functional profile is desired, the set of velocity-depth pairs may be fit to some function to result in a continuous depth profile. The fitting process may introduce additional parameters associated with the fit. The main drawback of the EDM is the necessary a priori assumption as to the functional form of the depth profile. In cases where the true profile does not resemble a linear or logarithmic function, the mapping results in errors in the velocities at the mapped depths. 

The EDM is illustrated in Figure \ref{fig:edm} using mock data with a small degree of noise artificially added, considering the horizontal velocity component of the Doppler shifts aligned with the current. Figure \ref{fig:edm}a) shows the Doppler shift velocities as a function of wavenumber (vertical axis) for the case of an exponential profile $U^{\mathrm{exp}}(z) = U_0\mathrm{e}^{z/d}$ with $U_0 = 1$ m/s and $d = 5$ m, and a linear profile $U^{\mathrm{lin}}(z) = U_0 + Sz$, with $S = 0.04$ $\mathrm{s}^{-1}$. The assumption of a linear profile, $Z_{\mathrm{eff}}(k) = -(2k)^{-1}$, is used to map the wavenumbers to depths, shown in Figure \ref{fig:edm}b). For comparison, the true current profiles are also shown. For the linear profile, the EDM maps the velocities to the correct depths as expected given a correct assumption concerning the functional form of the profile. In the case of the exponential profile, there is some deviation between the mapped currents and the true profile, since the mapping function is invalid in this case. Figure \ref{fig:edm}b highlights the main drawback of the EDM: the method leads to errors when the true profile differs from the assumed functional form used in the mapping.


\subsection{Ha-Campana method}
To avoid a priori assumptions of the functional form of the the current profile, Ha \cite{ha79} proposed a method which directly inverts the integral of (\ref{eq:SJ}) to find the unknown current profile in the integrand by approximating the integral using Gaussian quadrature. The method was later further developed and extended to finite depth using the integral of (\ref{eq:KC}) by Campana \textit{et al}. \cite{campana16}. We here outline the method for the case of infinite depth (the finite depth version is described in Reference \cite{campana16}).

An integral of a function $f(x)$ (assumed to be smooth) can be approximated as:

\begin{equation}
\int_{-1}^1f(x)\mathrm{d}x \approx \sum_{j=1}^nf(x_j)w_j,
\label{eq:gq}
\end{equation} 
where $x_j$ are quadrature points, $w_j$ are weights, and $n$ is the order of the Legendre polynomial. To match the integral limits of (\ref{eq:SJ}) to (\ref{eq:gq}), the coordinate transformation $x = 2\mathrm{e}^{-2k_0z}-1$ is made, where $k_0$ is a reference wavenumber chosen to minimize quadrature error\cite{ha79,campana17}.

For a set of Doppler shifts measured at discrete wavenumbers, (\ref{eq:SJ}) may be approximated using a matrix equation:

\begin{equation}
\mathbf{f} = \mathbf{A}\cdot\mathbf{u},
\label{eq:mateq}
\end{equation}
where column vector $\mathbf{f}$ contains the measured Doppler shifts, $\mathbf{A}$ is a matrix of coefficients derived from (\ref{eq:gq}), and $\mathbf{u}$ is a column vector with the unknown current velocities at discrete depths defined by the coordinate transformation (details given in References \cite{ha79} and \cite{campana17}.
%
The form of (\ref{eq:mateq}) is an over-determined linear system of equations, and may be solved in the least squares sense for the unknown values of the current $\mathbf{u}$ given a set of Doppler shift-wavenumber pairs. 

Up to this point, no assumptions concerning the functional form of the current profile have been made, a clear advantage relative to the EDM. However, the biggest challenge and drawback of the method concerns how errors in the Doppler shift velocities propagate through the inversion process to corresponding errors in the current depth-profile solution. Ideally, the resulting errors in the depth profile would be similar in magnitude to those of the input Doppler shifts. However, the direct inversion of the integrals (\ref{eq:SJ}) and (\ref{eq:KC}) result in a severe amplification of the error: small errors in the Doppler shift velocities result in large errors in the resulting depth-profile. The error amplification means that (\ref{eq:mateq}) is impractical to be used in reality without some additional constraints on the velocity solutions. The constraints typically imposed limit the second derivative of the velocity profile with respect to depth, and limit the distance from an initial guess. A cost function is typically defined and minimized, with increasing cost as the curvature or distance from the initial guess increases. Introduction of the constraints results in smoother profiles that suppress the amplification of the errors, yet the current profile solutions may depend on the values of the parameters weighting the constraints. The challenge then is how to choose optimal values of the constraint parameters. Campana \textit{et al}\cite{campana17} offer an empirical method for choosing the curvature constraint, though the universality of the method remains unclear. The Ha-Campana method demonstrates comparable accuracy relative to the EDM when compared to acoustic Doppler current profiler (ADCP) truth measurements, while reconstructing the current profile at a deeper range of depths.

\subsection{Polynomial effective depth method}
Another method for reconstructing the depth-profile without assumptions as to the functional form was proposed by Smeltzer \textit{et al}.\cite{smeltzer19}. The method starts from the conventional EDM, fits the profile to a polynomial form, and then scales the coefficients based on a simple-derived relation to produce an improved estimate of the true current profile. We outline the method below, considering one horizontal component of the Doppler shift velocity vector and current profile, expressed here as $\tilde{c}$ and $U(z)$ respectively. The method is in practice applied to both components separately. 

If we assume a polynomial form to the current profile, i.e. $U(z) = \sum_{n = 0}^\infty u_n z^n$, evaluation of the resulting Doppler shifts using (\ref{eq:SJ}) yields
\begin{equation}
  \tilde{c}(k) = \sum_{n = 0}^\infty n!u_n \left(-\frac{1}{2k}\right)^n.
\label{eq:pedm_ctil}
\end{equation}
By inspecting (\ref{eq:pedm_ctil}) we notice that the $(-2k)^{-1}$-term is equal to the mapping function $Z_{\mathrm{eff}}(k)$ of the EDM for a linear profile, (\ref{eq:edmlin}). If we substitute the EDM mapping function, we then obtain the current profile
\begin{equation}
  U_{\mathrm{EDM}}(z) = \sum_{n = 0}^\infty n!u_n z^n.
\label{eq:pedm}
\end{equation}
We see that $U_{\mathrm{EDM}}(z)$ differs from the true profile $U(z)$ only by a factor $n!$ for the $n$-th order term. The $n!$ characterizes the error of the EDM profile in cases where the linear assumption is invalid. For term $n<2$, (\ref{eq:pedm}) matches the true profile as expected, while differing for higher order terms which represent a profile with nonlinear functional form. The similarity of (\ref{eq:pedm}) to the form of the true profile motivates the method described here, the polynomial effective depth method (PEDM), which attempts to correct for the discrepancy in terms $n\geq 2$ by simply scaling the higher order polynomial coefficients by a factor $n!$. Quoted from Reference \cite{smeltzer19}, the PEDM procedure consists of three steps:

\begin{enumerate}
  \item For each of the measured values $\tilde{c}_i$, assign effective depths $z_i = -(2k_i)^{-1}$ according to the EDM procedure of (\ref{eq:edmlin}) using $Z_{\mathrm{eff}}(k)$.
  \item Obtain $U_{\mathrm{EDM}}(z)$ by fitting the set of points $\{z_i,\tilde{c}_i\}$ to a polynomial of degree $n_\mathrm{max}$:
    \begin{equation}\label{eq:EDMfit}
      U_{\mathrm{EDM}}(z)\approx\sum_{n=0}^{n_\mathrm{max}} u_{\mathrm{EDM},n}z^n,
    \end{equation}
    where $u_{\mathrm{EDM},n}$ are the coefficients obtained in the polynomial fit.
  \item Then the improved PEDM estimate is 
    \begin{equation}\label{eq:PEDM2}
      U_{\mathrm{PEDM}}(z)=\sum_{n=0}^{n_\mathrm{max}} \frac1{n!} u_{\mathrm{EDM},n}z^n.
    \end{equation}
    
\end{enumerate}
Equation (\ref{eq:PEDM2}) follows immediately from comparing (\ref{eq:pedm}) and (\ref{eq:EDMfit}), where $u_{\mathrm{EDM},n} = n!u_n$. Additional practical details concerning the implementation of the method are given in Reference \cite{smeltzer19}.

\begin{figure}[h]
\centering \includegraphics[scale=0.8]{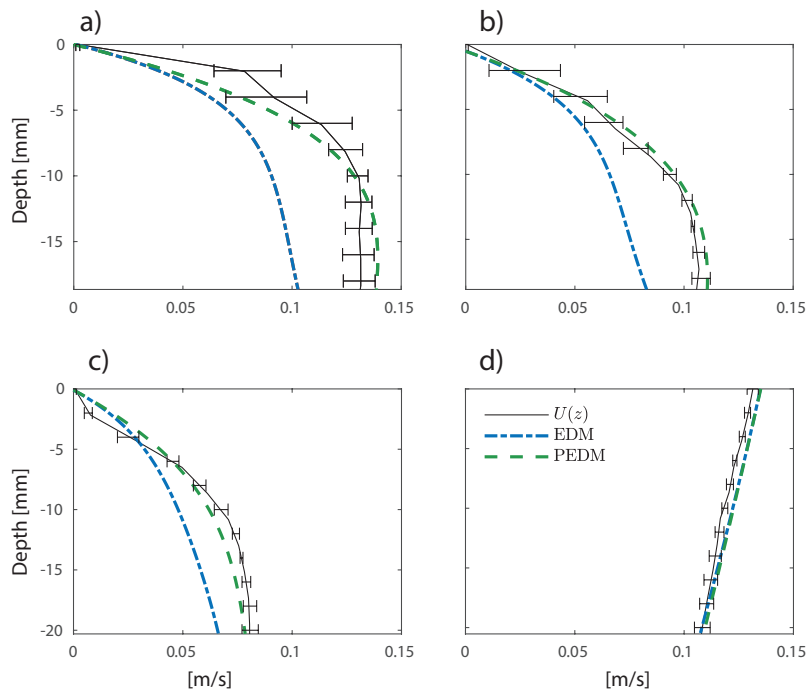}
\caption{Data from Reference \cite{myRep} (presented in Reference \cite{smeltzer19}). A comparison of the PEDM and EDM methods applied to experimentally-measured Doppler shifts along the direction of the flow for four different current profiles in panels a-d). In situ truth measurements are denoted as $U(z)$. }
\label{fig:pedm}
\end{figure}

The PEDM has been tested on Doppler shifts measured in a laboratory with currents of variable depth dependence, with in situ particle image velocimetry serving as ``truth'' measurements. The laboratory setup may be considered a scale model of the oceanographic currents, including wind-drift profiles. The wave spectrum was measured by optical means, yet representative of what may be obtained by X-band radar for scaled-down length dimensions. Example results are shown in Figure \ref{fig:pedm} for four different current profiles. Using the PEDM resulted in a $>$3 times accuracy improvement relative to the EDM for current profiles with significant near surface curvature (profiles a-c). For one profile that varied approximately linearly with depth (profile d), the PEDM and EDM resulted in essentially indistinguishable profiles as expected since in that case the assumptions inherent to the EDM were fulfilled. 

%
\section{Challenges and further work}
So far in this chapter we have examined different methods for extracting currents from the wave spectrum, both assuming a depth-uniform as well as depth-varying profile. We now outline some challenges associated with evaluating and interpreting the results, which may be the focus of future efforts within the field.
\subsection{Validation}
\begin{figure}[h]
\centering \includegraphics[scale=0.65]{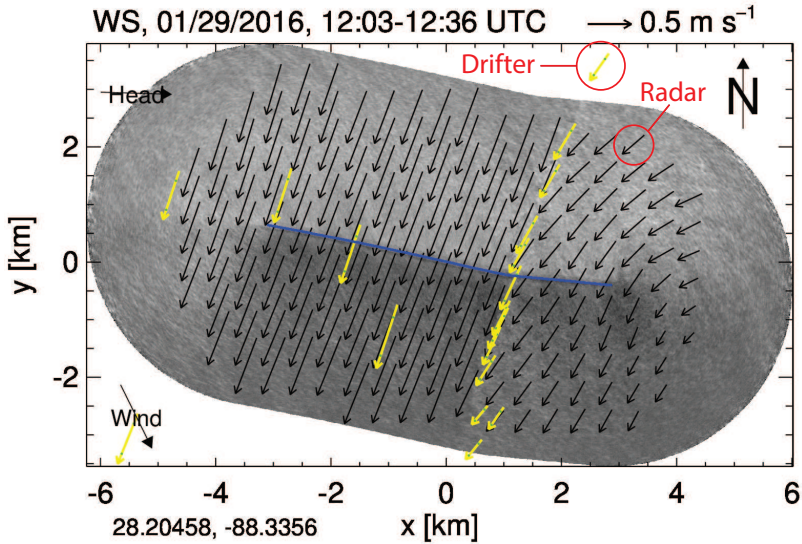}
\caption{Reproduced from \cite{lundRep}, a comparison of radar and drifter currents. } 
\label{fig:lund18}
\end{figure}

A key area of work within X-band current mapping concerns the validation of the methods described previously in this chapter, and evaluation of their accuracy. Of particular interest is the absolute accuracy, as well as identifying what factors affect the accuracy and reliability of reconstructed current maps. Comparison is typically performed relative to in situ measurements such as ADCP or drifters. Field studies, such as Reference \cite{lund18} with selected data in Figure \ref{fig:lund18} have demonstrated agreement down to nearly cm/s-scale between radar-derived and in situ measurements assuming a depth-uniform flow. 

One challenge when comparing to in situ measurements is due to different sampling of the current field between the two approaches: the currents obtained from radar are representative of the current over a finite areal extent in the horizontal plane whereas in situ measurements are point measurements. In areas with strong horizontal current shear or local variations within the radar spatial window, in situ results may give different results simply because they sample only a single discrete point in the horizontal plane whereas radar-derived currents are representative of a spatial average of the currents of the window extent. Point measurements at many different locations may be achieved conveniently using drifters, though they have the disadvantage of tending to congregate at convergent zones \cite{lund18} and thus potentially not evenly sample the horizontal area. An illustrative example is shown in Figure \ref{fig:lund18} where the drifters (marked as the yellow arrows with green dots), clearly congregate near a convergent zone of the currents.
 
Another challenge when comparing to some in situ measurements concerns the fact that radar derived currents may include Lagrangian components (following the movement of a particular fluid parcel) such as waves Stokes drift whereas some in situ techniques such as ADCP measure the currents in an Eulerian framework (the velocity at a fixed point in space). We elaborate on Stokes drift in the next subsection. It is noted that drifters also include Lagrangian components and may offer a more direct comparison in cases where such Lagrangian current components are relevant.

In addition, ADCP may provide measurements of the depth profile, but many in situ techniques measure the current at a single depth. When using a depth-uniform assumption, the comparison is thus more complicated since the radar-derived currents represent a weighted average over a depth range determined by the wavenumbers as discussed in section \ref{sec:depth}. Discrepancies between radar and in situ measurements may in many cases be due to the different depths to which the currents correspond. A few field studies have reconstructed the depth profile using one of the methods described in section \ref{sec:depth} and compared to in situ measurements. In general, more validation is required for testing the accuracy of the methods in section \ref{sec:depth}. A particular challenge concerns reliable truth measurements of the depth profile in the upper meters of the water column, a relative ``blind spot'' of current sensing technology. Currents reconstructed from waves are an attractive method in the near-surface regime which introduces a paradox of sorts: radar-derived currents are attractive since they can provide measurements in this blind spot where few other reliable methods exist, yet this in turn makes validation a challenge because truth measurements are difficult to obtain.

\subsection{Interpretation of the currents: Stokes drift}
As mentioned above, radar-derived currents may include Lagrangian components, namely the waves-induced Stokes drift. A fluid parcel considered in a Lagrangian framework perturbed by waves follows an oscillatory trajectory. In addition to the oscillation, there is a net translation of the parcel in the direction of the wave (expect in the limit of infinitesimal wave amplitude) which is known as Stokes drift. It has been proposed that the currents measured by radar systems are a sum of a background Eulerian currents (such as a wind drift of tidal current) and a Stokes drift component \cite{lund15}:

\begin{equation}
\mathbf{U}_R = \mathbf{U}_{E} + \mathbf{U}_{SS},
\end{equation}
where $\mathbf{U}_{E}$ and $\mathbf{U}_{SS}$ are the Eulerian and surface Stokes drift components respectively and $\mathbf{U}_{R}$ is the total current measured from the wave spectrum. The surface Stokes drift component is a function of the wave energy spectrum reflecting the heights of the waves, and has been suggested to be expressed as \cite{ardhuin09}

\begin{equation}
\mathbf{U}_{SS} = 4\pi\int_0^{f_B}\int_0^{2\pi}f\mathbf{k}(f)E(f,\theta)\mathrm{d}f\mathrm{d}\theta,
\label{eq:stokes}
\end{equation}
where frequency $f = \omega/2\pi$, $E(f,\theta)$ is the wave energy density spectrum as a function of frequency and direction $\theta$, $\mathbf{k}(f)$ is an alternate form of the dispersion relation, and  $f_B$ is the frequency of the Bragg-resonant wave determined by the radar wavelength. The Stokes drift component decays rapidly with depth, complicating the interpretation of $\mathbf{U}_R$ at greater depths. Given measurements of $E(f,\theta)$, the Stokes drift component may be estimated and subtracted using (\ref{eq:stokes}), allowing separate analysis of the Eulerian current components \cite{lund15}). It has been found that the Stokes drift under relevant conditions may be on the order of 5-10 cm/s as discussed in Reference \cite{lund20}, non-negligible relative to the current strengths typically derived from radar measurements.

However, there remains ongoing debate about the magnitude of the surface Stokes drift for a spectrum of waves (i.e. the validity of (\ref{eq:stokes})), and the extent to which Stokes drift is a component of radar-derived currents \cite{rohrs15,chavanne18}. Different theoretical formulations have been proposed, as summarized in Reference \cite{chavanne18}. Further research is required on the matter, which is clearly important to an increased understanding of the physical interpretation of the currents measured from radar images.

\section{Summary}
In this chapter we have examined how spatially-varying ocean currents can be extracted from remote measurements of the wave spectrum. For marine radar images, the spatial variation of currents may be mapped within the full radar field of regard, an attractive means of current remote sensing that has multiple advantages to in situ point sensors. Currents are extracted by analyzing their effect on wave propagation, appearing as frequency-dependent shifts in the linear dispersion relation curve along which the wave spectrum is strongly peaked. Several algorithms were described for obtaining empirical dispersion relations from the measured spectrum and extracting the currents: the least squares and iterative least squares method, the normalized scalar product method, and the polar current shell method. 

We go on to describe how the same methods and algorithms can be extended to also allowing the depth-dependence of the current to be determined. Multiple velocities must now be extracted over different wavenumber bins, whereupon the set of velocities at varying wavenumber is further analyzed using an inversion method to find the depth dependence.

Reasonable agreement between radar-derived currents and in situ measurements has been demonstrated in multiple field measurements. However, more validation is necessary especially in the context of depth-varying flows where comparison is more difficult and in situ data is scarce. Understanding the extent to which the Lagrangian current from the waves, Stokes drift, is measured as part of the radar-derived current is not well-understood, yet important for interpretation of the mapped currents and comparison with in situ measurements.


\bibliographystyle{unsrt}


\begin{thebibliography}{10}

\bibitem{kudryavtsev08}
V.~Kudryavtsev, V.~Shrira, V.~Dulov, and V.~Malinovsky.
\newblock On the vertical structure of wind-driven sea currents.
\newblock {\em J.~Phys.\ Ocean.}, 38:2121--44, 2008.

\bibitem{belcher12}
S.E. Belcher, A.~L.~M. Grant, K.~E. Hanley, B.~Fox-Kemper, L.~V. Van~Roekel,
  and P.~\textit{et al}. Sullivan.
\newblock A global perspective on {L}angmuir turbulence in the ocean surface
  boundary layer.
\newblock {\em Geophys.\ Res.\ Lett.}, 39:L18605, 2012.

\bibitem{laxague17}
N.~J.~M. Laxague, B.~K. Haus, D.~G. Ortiz-Suslow, C.~J. Smith, G~Novelli, and
  H~\textit{et al}. Dai.
\newblock Passive optical sensing of the near-surface wind-driven current
  profile.
\newblock {\em J.~Atmos.\ Ocean.\ Technol.}, 34:1097--111, 2017.

\bibitem{laxague18}
N.~J.~M. Laxague, T.~M. \"{O}zg\"{o}kmen, B.~K. Haus, G.~Novelli,
  A.~Shcherbina, and P~\textit{et al}. Sutherland.
\newblock Observations of near-surface current shear help describe oceanic oil
  and plastic transport.
\newblock {\em Geophys.\ Res.\ Lett.}, 45:245249, 2018.

\bibitem{lund18}
B.~Lund, B.K. Haus, J.~Horstmann, H.~C. Graber, R.~Carrasco, and N.~J.
  M.~\textit{et al}. Laxague.
\newblock Near-surface current mapping by shipboard marine {X}-band radar: a
  validation.
\newblock {\em J.~Atmos.\ Ocean.\ Technol.}, 35:1077--90, 2018.

\bibitem{gangeskar18}
R.~Gangeskar.
\newblock Verifying high-accuracy ocean surface current measurements by
  {X}-band radar for fixed and moving installations.
\newblock {\em IEEE Trans.\ Geosci.\ Remote Sens.}, 56(8):4845--55, 2018.

\bibitem{zippel17}
S.~Zippel and J.~Thomson.
\newblock Surface wave breaking over sheared currents: Observations from the
  mouth of the {C}olumbia {R}iver.
\newblock {\em J.~Geophys.\ Res.: Oceans}, 122(4):3311--28, 2017.

\bibitem{dalyrmple73}
R.~A. Dalrymple.
\newblock Water wave models and wave forces with shear currents.
\newblock Technical Report no. 20, Coastal and Oceanographic Engineering
  Laboratory, Uni.\ Florida, Gainesville (FL), 1973.

\bibitem{young85}
I.~R. Young and W.~Rosenthal.
\newblock A three-dimensional analysis of marine radar images for the
  determination of ocean wave directionality and surface currents.
\newblock {\em J.~Geophys.\ Res.}, 90:1049--59, 1985.

\bibitem{lund15}
B.~Lund, H.~C. Graber, H.~Tamura, C.~O. Collins~III, , and S.~M. Varlamov.
\newblock A new technique for the retrieval of near-surface vertical current
  shear from marine {X}-band radar images.
\newblock {\em J.~Geophys.\ Res.: Oceans}, 120:8466--86, 2015.

\bibitem{campana17}
J.~Campana, E.~J. Terrill, and T.~de~Paolo.
\newblock A new inversion method to obtain upper--ocean current--depth profiles
  using {X}-band observations of deep--water waves.
\newblock {\em J.~Atmos.\ Oceanic Technol.}, 34:957--70, 2017.

\bibitem{hessner14}
K.~Hessner, K.~Reichert, J.~C.~N. Borge, C.~L. Stevens, and M.~J. Smith.
\newblock High-resolution {X}-band radar measurements of currents, bathymetry
  and sea state in highly inhomogeneous coastal areas.
\newblock {\em Ocean Dyn.}, 64:989--998, 2014.

\bibitem{lundRep}
B.~Lund and B.~Haus.
\newblock Radar measurements collected during the {L}agrangian {S}ubmesoscale
  {E}xpe{R}iment ({LASER}) experiment aboard {R}/{V} {W}alton {S}mith cruise
  {WS}16015 in the {G}ulf of {M}exico from 2016-01-20 to 2016-02-12.
  {D}istributed by: {G}ulf of {M}exico {R}esearch {I}nitiative {I}nformation
  and {D}ata {C}ooperative ({GRIIDC}), {H}arte {R}esearch {I}nstitute, {T}exas
  {A}\&{M} {U}niversity -- {C}orpus {C}hristi., 2018.
\newblock Online data set available from http://doi.org/10.7266/N7N01550.

\bibitem{horstmann17}
J.~Horstmann, M.~Stresser, and R.~Carrasco.
\newblock Surface currents retrieved from airborne video.
\newblock In {\em Proceedings OCEANS 2017-Aberdeen}, 2017.

\bibitem{micha17}
M.~Stre{\ss}er, R.~Carrasco, and J.~Horstmann.
\newblock Video-based estimation of surface currents using a low-cost
  quadcopter.
\newblock {\em IEEE Geosci.\ Remote Sens.\ Lett.}, 14(11):2027--31, 2017.

\bibitem{lund15a}
B.~Lund, H.~C. Graber, K.~Hessner, and N.~J. Williams.
\newblock On shipboard marine {X}-band radar near-surface current
  ``calibration''.
\newblock {\em J.~Atmos.\ Ocean.\ Technol.}, 32:1928--1944, 2015.

\bibitem{mccann18}
D.~L. Mc{C}ann and P.~S. Bell.
\newblock A simple offset ``calibration'' method for the accurate geographic
  registration of ship-borne {X}-band radar intensity imagery.
\newblock {\em IEEE Access.}, 6:13939--13948, 2018.

\bibitem{senet01}
C.~M. Senet, J.~Seeman, and F.~Ziemer.
\newblock The near-surface current velocity determined from image sequences of
  the sea surface.
\newblock {\em IEEE Trans.\ Geosci.\ Remote Sens.}, 39:492--505, 2001.

\bibitem{huang12}
W.~Huang and E.~Gill.
\newblock Surface current measurement under low sea state using dual polarized
  {X}-band nautical radar.
\newblock {\em IEEE J.\ Sel.\ Topics Appl.\ Earth Observ.\ Remote Sens.},
  5:1868--73, 2012.

\bibitem{gangeskar02}
R.~Gangeskar.
\newblock Ocean current estimated from {X}-band radar sea surface images.
\newblock {\em IEEE Trans.\ Geosci.\ Remote Sens.}, 40:783--92, 2002.

\bibitem{serafino10}
F.~Serafino, C.~Lugni, and F.~Soldovieri.
\newblock A novel strategy for the surface current determination from marine
  {X}-band radar data.
\newblock {\em IEEE Geosci.\ Remote Sens.\ Lett.}, 7:231--35, 2010.

\bibitem{huang16}
W.~Huang, R.~Carrasco, C.~Shen, E.~W. Gill, and J.~Horstmann.
\newblock Surface current measurements using {X}-band marine radar with
  vertical polarization.
\newblock {\em IEEE Trans.\ Geosci.\ Remote Sens.}, 54:2988--97, 2016.

\bibitem{shen15}
C.~Shen, W.~Huang, E.~Gill, R.~Carrasco, and J.~Horstmann.
\newblock An algorithm for surface current retrieval from {X}-band marine radar
  images.
\newblock {\em Remote Sens.}, 7:7753--67, 2015.

\bibitem{smeltzer19}
B.~K. Smeltzer, E.~{\AE}s{\o}y, A~{\AA}dn{\o}y, and S.~\AA. Ellingsen.
\newblock An improved method for determining near-surface currents from wave
  dispersion measurements.
\newblock {\em J.~Geophys.\ Res.: Oceans}, 124:8832â€“8851, 2019.

\bibitem{stewart74}
R.~J. Stewart and J.~W Joy.
\newblock {HF} radio measurements of surface currents.
\newblock {\em Deep Sear Res. Oceanograph. Abs.}, 21:1039--49, 1974.

\bibitem{skop87}
R.~A. Skop.
\newblock Approximate dispersion relation for wave-current interactions.
\newblock {\em J. Waterway, Port, Coastal, and Ocean Eng.}, 113:187--95, 1987.

\bibitem{kirby89}
J.~T. Kirby and T.-M. Chen.
\newblock Surface waves on vertically sheared flows: approximate dispersion
  relations.
\newblock {\em J. Geophys. Res.}, 94:1013--27, 1989.

\bibitem{lund20}
B.~Lund, B.K. Haus, H.~C. Graber, J.~Horstmann, R~Carrasco, and G.~\textit{et
  al}. Novelli.
\newblock Marine {X}-band radar currents and bathymetry: An argument for a wave
  number-dependent retrieval method.
\newblock {\em J.~Geophys.\ Res.: Oceans}, 125:e2019JC015618, 2020.

\bibitem{wu75}
J.~Wu.
\newblock Wind-induced drift currents.
\newblock {\em J. Fluid Mech.}, 68:49--70, 1975.

\bibitem{plant80}
W.J. Plant and J.~W. Wright.
\newblock Phase speeds of upwind and downwind traveling short gravity waves.
\newblock {\em J.~Geophys.\ Res.}, 85:3304--10, 1980.

\bibitem{fernandez96}
D.M. Fernandez, J.F. Vesecky, and C.~Teague.
\newblock Measurememts of upper ocean surface current shear with high-frequency
  radar.
\newblock {\em J.~Geophys.\ Res.}, 101:28615--25, 1996.

\bibitem{teague01}
C.~C. Teague, J.~F. Vescky, and Z.~R. Hallock.
\newblock A comparison of multifrequency {HF} radar and {ADCP} measurements of
  near-surface currents during cope-3.
\newblock {\em IEEE J.~Oceanic Eng.}, 26:399--405, 2001.

\bibitem{ha79}
E.-C. Ha.
\newblock Remote sensing of ocean surface current and current shear by hf
  backscatter radar.
\newblock Technical Report Technical Report No. D415-1, Stanford Uni., Stanford
  (CA), 1979.

\bibitem{campana16}
J.~Campana, E.~J. Terrill, and T.~de~Paolo.
\newblock The development of an inversion technique to extract vertical current
  profiles from {X}-band radar observations.
\newblock {\em J.~Atmos.\ Oceanic Technol.}, 33:2015--28, 2016.

\bibitem{myRep}
B.~K. Smeltzer.
\newblock Replication data for: New method to determine near-surface currents
  from measurements of the wave spectrum, {D}ataverse{NO}, {V}1., 2019.
\newblock Online data set available from https://doi.org/10.18710/8JBWCJ.

\bibitem{ardhuin09}
F.~Ardhuin, L.~Mari\'{e}, N.~Rascle, P.~Forget, and A.~Roland.
\newblock Observation and estimation of {L}agrangian, {S}tokes, and {E}ulerian
  currents induced by wind and waves at the sea surface.
\newblock {\em J.\ Phys. Oceanogr.}, 39:2820--38, 2009.

\bibitem{rohrs15}
J.~R\"{o}hrs, A.~Sperrevik, K.~Christensen, \O. Breivik, and G.~Brostr\"{o}m.
\newblock Comparison of {HF} radar measurements with eulerian and lagrangian
  surface currents.
\newblock {\em Ocean Dyn.}, 65:679--90, 2015.

\bibitem{chavanne18}
C.~Chavanne.
\newblock Do {H}igh-{F}requency radars measure the wave-induced {S}tokes drift?
\newblock {\em J.~Atmos.\ Oceanic Technol.}, 35:1023--31, 2018.

\end{thebibliography}

\end{document}